\documentclass[10pt]{article}
\usepackage[margin=0.78in]{geometry}
\usepackage[T1]{fontenc}
\usepackage{lmodern}
\usepackage{microtype}
\usepackage{amsmath,amssymb}
\usepackage{booktabs,tabularx,array,multirow,makecell}
\usepackage{enumitem}
\usepackage{graphicx}
\usepackage{float}
\usepackage{caption}
\usepackage{subcaption}
\usepackage{xcolor}
\usepackage{tikz}
\usetikzlibrary{arrows.meta,positioning,shapes.geometric,fit,calc,backgrounds}
\usepackage{hyperref}
\usepackage{url}
\usepackage{xurl}
\usepackage{fancyhdr}
\usepackage{titlesec}
\usepackage{ragged2e}
\usepackage{longtable}
\usepackage{ltablex}
\keepXColumns

\definecolor{navy}{RGB}{31,65,104}
\definecolor{teal}{RGB}{39,116,118}
\definecolor{orange}{RGB}{196,116,48}
\definecolor{redsoft}{RGB}{165,70,70}
\definecolor{green}{RGB}{70,125,80}
\definecolor{graybox}{RGB}{242,244,247}
\definecolor{lightblue}{RGB}{232,240,248}
\definecolor{lightgreen}{RGB}{236,246,238}
\definecolor{lightorange}{RGB}{250,242,231}
\definecolor{lightred}{RGB}{249,235,235}

\hypersetup{colorlinks=true,linkcolor=navy,citecolor=navy,urlcolor=navy,pdftitle={Beyond Code Generation: Reliability, Verification, and Cost Economics in the Agentic Software Development Lifecycle},pdfauthor={Happy Bhati}}
\setlist[itemize]{leftmargin=1.3em,itemsep=2pt,topsep=2pt}
\setlist[enumerate]{leftmargin=1.5em,itemsep=2pt,topsep=2pt}

\titleformat{\section}{\large\bfseries\color{navy}}{\thesection}{0.5em}{}
\titleformat{\subsection}{\normalsize\bfseries\color{navy}}{\thesubsection}{0.45em}{}
\titleformat{\subsubsection}{\normalsize\bfseries}{\thesubsubsection}{0.4em}{}

\newcommand{\pqc}{\mathrm{PQC}}

\begin{document}

\begin{center}
{\LARGE\bfseries Beyond Code Generation: Reliability, Verification, and Cost Economics in the Agentic Software Development Lifecycle}\par
\vspace{5pt}
{\large A Systems Synthesis of Industrial Evidence and a Research Agenda for Agentic Engineering}\par
\vspace{10pt}
{\large Happy Bhati}\par
{\small \href{mailto:bhati.h@northeastern.edu}{bhati.h@northeastern.edu}}\par
\vspace{4pt}
{\small September 2026}
\end{center}

\begin{abstract}
AI coding systems are moving from autocomplete and chat toward agents that can inspect repositories, edit multiple files, run tools, write tests, open pull requests, and work for long periods with limited supervision. That capability changes the bottleneck in software delivery. Recent field experiments show large gains in coding activity, but newer evidence also shows that those gains attenuate sharply between writing code and shipping reliable software. Review, integration, testing, security, deployment, and production operations remain constraining stages; in several studies, the quality of the surrounding harness and verification process matters as much as raw model capability. At the same time, the economics are changing from predictable per-seat licenses toward variable token, tool, sandbox, CI, and rework costs.

This paper synthesizes peer-reviewed software-engineering research, university studies, benchmark audits, production reports from major technology companies, developer telemetry, and cost-management evidence released primarily from 2024 through September 2026. It does not report new model experiments. Numerical findings remain attributed to their original studies. The synthesis develops four engineering concepts: the Agentic SDLC Throughput Paradox, which describes why code-generation gains can outrun release gains; Production-Qualified Change (PQC), a unit of output that counts a change only after reliability gates are satisfied; the Verification Tax, which makes downstream review and assurance costs explicit; and an Agentic SDLC Control Plane that allocates autonomy subject to cost, reliability, and human-attention budgets. The paper then maps an evidence-based horizon from today's supervised coding agents to future policy-bounded software factories. The resulting research agenda argues that the useful question is no longer how much code an agent can generate, but how much production-qualified value an engineering system can deliver per dollar, per reviewer-hour, and per unit of operational risk.
\end{abstract}


\section{Introduction}

Software engineering is entering an odd phase: producing a plausible patch is getting cheaper at the same time that proving the patch deserves to ship is becoming more important. Coding agents can now search a repository, modify code, run commands, and iterate on failures. This is a genuine capability change, not merely a faster autocomplete. My earlier work described this transition as an emerging Agentic Software Development Lifecycle (A-SDLC) and focused on architecture, delegated execution, and changes in engineering work \cite{bhati2026agentic}. The present paper starts one step later. It asks what happens when organizations try to operate that lifecycle at scale.

The answer emerging from recent evidence is not simply ``developers become faster.'' In three randomized field experiments spanning 4,867 developers, AI assistance increased completed tasks by 26.08\% in the pooled estimate \cite{cui2026}. That is strong causal evidence that assistance can improve output in some professional settings. Yet a 2026 study of more than 100,000 GitHub developers found a much larger cumulative increase in coding activity from autonomous agents - 180\% at the commit level - that fell to 50\% at the project level and 30\% at actual releases \cite{demirer2026}. The authors describe a weak-link production structure: writing code accelerates faster than the human and organizational stages needed to turn changes into shipped software.

That gap is visible from several directions. Google reports millions of code-review comments annually and about 60 minutes of active author shepherding time between sending a change for review and submitting it \cite{frommgen2024}. Google's 2025 DORA study, based on nearly 5,000 technology professionals, associated higher AI adoption with higher delivery throughput but still found a negative relationship with delivery stability \cite{dora2025}. Stanford's SWE-chat dataset of 6,000 real coding-agent sessions found that only 44\% of agent-produced code survived into user commits, while users corrected, interrupted, or otherwise pushed back on agent outputs in 44\% of turns; agent-written code also showed more security vulnerabilities than human-written code in that dataset \cite{baumann2026}. Long-horizon evaluations reveal a different failure shape: SWE-Marathon rollouts averaged 27.2 million tokens, yet the initial study found no tested configuration above 30\% pass@1 and observed reward-hacking behavior in 13.8\% of rollouts \cite{desai2026}.

These are not contradictory results. They describe different layers of the same system. AI can increase local coding productivity while the global delivery system remains constrained by verification, coordination, human judgment, and operational risk. In an agentic workflow, those constraints are not side issues. They become part of the runtime.

Cost follows the same pattern. The direct model bill is easy to see. The rest is distributed across context construction, repeated attempts, tool calls, sandboxes, CI minutes, security scans, reviewer attention, rework, incidents, and the infrastructure required to observe all of it. Northeastern researchers show that token use can vary substantially across programming languages even after controlling for problem difficulty, in part because agents produce non-compiling attempts, revise already-passing solutions, and distrust provided tests \cite{wu2026}. The FinOps Foundation reports that 98\% of surveyed practitioners now manage AI spend, up from 31\% two years earlier, and identifies AI cost management as the top skillset to develop \cite{finops2026}. Gartner goes further and forecasts that AI coding cost could exceed the average developer salary by 2028 under growing token consumption and consumption-based pricing \cite{gartner2026codingcost}. That forecast should not be treated as a measured inevitability, but it is a useful signal that engineering leaders are moving from adoption questions to unit-economics questions.

This paper connects those threads. It treats an AI coding agent as one component in a production system, not as the unit of evaluation. The goal is to identify the control problem organizations now face: how much autonomy should a task receive, which model and harness should execute it, how much verification is required, when should a human intervene, and when does another agent attempt stop being worth its cost?

The synthesis is organized around four concepts:
\begin{enumerate}
    \item \textbf{Agentic SDLC Throughput Paradox.} Upstream change generation can accelerate much faster than downstream validation and release, so local productivity growth does not translate linearly into shipped value.
    \item \textbf{Production-Qualified Change (PQC).} A pull request, commit, or generated patch is not finished output. A change becomes useful production output only after it crosses the relevant review, test, security, deployment, and operational gates.
    \item \textbf{Verification Tax.} Agentic generation creates a variable assurance workload. Its cost includes CI, reviewer time, security analysis, rework, and escaped failures, not only tokens.
    \item \textbf{Agentic SDLC Control Plane.} A policy and telemetry layer should allocate models, context, parallelism, retries, tests, and human review according to task risk, reliability evidence, budget, and organizational capacity.
\end{enumerate}

The purpose is not to claim that any one company has already implemented this complete control plane. The purpose is to turn a scattered set of research results into a systems model that can be tested.

\section{Scope, Evidence Selection, and Claim Boundary}

The review emphasizes work from 2024 through September 3, 2026, with earlier benchmark papers included where they establish an important baseline. Sources were selected for direct relevance to at least one of six questions: (1) coding-agent capability, (2) developer productivity and work allocation, (3) code review and testing, (4) reliability and security, (5) coordination and long-horizon execution, or (6) cost and resource governance.

The source set intentionally mixes research types because production software engineering is not well described by benchmark papers alone. A randomized field experiment answers a different question from a public leaderboard; an industrial deployment report answers a different question from a controlled academic study; a market forecast is weaker evidence than either, but can still describe what organizations are budgeting for. Table~\ref{tab:evidence} keeps those categories visible.

\begin{table}[H]
\centering
\caption{Evidence hierarchy used in this synthesis. Tiers describe evidence type, not a universal ranking of research quality.}
\label{tab:evidence}
\small
\begin{tabularx}{\textwidth}{p{0.8cm} p{3.0cm} X X}
\toprule
\textbf{Tier} & \textbf{Evidence type} & \textbf{Representative sources} & \textbf{How claims are used} \\
\midrule
A & Peer-reviewed / major conference research & Management Science developer RCTs; Google ICSE code-review work; Meta FSE TestGen-LLM; Microsoft ASE developer-agent study & Strong basis for findings in the studied setting; not generalized beyond the design. \\
B & Academic preprints / university research & Stanford SWE-chat and CooperBench; SWE-Marathon; MIT/NBER shipping study; Northeastern token-cost study & Frontier evidence; useful for identifying emerging failure modes and hypotheses. \\
C & Benchmark / evaluation research & SWE-bench, SWE-Lancer, benchmark audits, long-horizon benchmarks & Used to measure specific capabilities; benchmark limitations are treated as part of the result. \\
D & Documented industrial telemetry / deployment & Google DORA; GitHub security validation and ROI dashboard; company telemetry and engineering reports & Evidence of practice or observed telemetry, with vendor/self-report boundary stated. \\
E & Cost / market / workforce reports & FinOps Foundation; Gartner forecasts & Used for adoption and planning signals; forecasts are explicitly not treated as measured outcomes. \\
F & This paper's synthesis & PQC, Verification Tax, Autonomy Budget, Control Plane, horizon map & Proposed engineering constructs to be validated, not empirical results. \\
\bottomrule
\end{tabularx}
\end{table}

A second boundary is important because this paper builds on my prior A-SDLC paper \cite{bhati2026agentic}. The earlier paper mapped the architectural transition from assistant to agent. This paper does not repeat that architecture as a new contribution. It narrows the question to reliability, verification capacity, delivery economics, and the operating model required when agentic work becomes routine. One additional self-citation is used for AI observability, where prior work proposed tracing model cost and quality together \cite{bhati2026observability}. The aim is continuity, not citation inflation.

No benchmark score, productivity gain, code-review statistic, security result, market number, or company deployment outcome in this paper is an original measurement. Those results remain with the cited authors and organizations.

\section{From Coding Capability to Production Capability}

\subsection{Benchmarks established the capability leap}

SWE-bench moved coding evaluation from small algorithmic problems toward real GitHub issues and repository states \cite{jimenez2024}. SWE-agent showed that the agent-computer interface around a model materially affects performance, making tools and interaction design part of the system rather than neutral plumbing \cite{yang2024sweagent}. OpenAI's SWE-Lancer added an economic dimension by collecting more than 1,400 real freelance software-engineering tasks representing roughly US\$1 million in historical payouts \cite{miserendino2025}. These benchmarks made agent progress legible and pushed models toward repository-scale work.

The benchmarks also exposed a problem that becomes more important as scores rise: automated tests are not a perfect proxy for software correctness. OpenAI's 2026 audit of SWE-bench Verified reported that, among 138 tasks that one frontier model did not consistently solve, 59.4\% had material issues in test design or problem description. The same analysis identified contamination risk from public tasks and solutions \cite{openai2026swebench}. A separate position paper argues that coding benchmarks conflate the model, harness, context, execution environment, and feedback signals into a single score, making it difficult to know what actually improved \cite{gorinova2026}.

For production engineering this matters because a high benchmark score is evidence about one slice of the delivery path. It is not a release-readiness certificate. A benchmark usually starts with a well-formed task and ends when a hidden test suite passes. A company starts earlier - with incomplete requirements and organizational context - and ends later, after security review, deployment, monitoring, user behavior, and maintenance.

\subsection{Software engineering is broader than patch generation}

Gu et al., with authors from MIT, UC Berkeley, Stanford, and Cornell, argue that current AI-for-software-engineering work over-concentrates on code generation relative to the breadth of real engineering activity: testing, debugging, review, refactoring, performance, security, migration, communication, and large-scale change management all matter \cite{gu2025}. Microsoft Research reaches a similar conclusion from direct observation. In a 2025 ASE study of 19 developers resolving 33 real repository issues with an interactive agent, roughly half the issues were successfully resolved; incremental collaboration outperformed one-shot use, and developers still struggled with trust, debugging, and testing \cite{kumar2025}.

The practical implication is that ``agent capability'' should be decomposed. A useful organization-level question is not whether an agent can write a patch, but whether the entire human-agent system can consistently move a change from intent to production without creating more validation work than it removes.

\section{The Agentic SDLC Throughput Paradox}

The clearest recent evidence for a bottleneck migration comes from Demirer, Musolff, and Yang \cite{demirer2026}. Using data from more than 100,000 GitHub developers combined with AI usage telemetry, they estimate progressively larger cumulative effects on commits as tools move from autocomplete to interactive and autonomous agents: 40\%, 140\%, and 180\%, respectively. But at the autonomous-agent stage, the effect is only 50\% for the number of projects and 30\% for releases.

\begin{figure}[H]
\centering
\begin{tikzpicture}[x=0.055cm,y=1cm,font=\small]
\node[anchor=west,font=\bfseries] at (0,3.55) {Cumulative effect after autonomous-agent adoption};
\fill[navy!78] (0,2.55) rectangle (180,3.10);
\node[anchor=east,text=white,font=\bfseries] at (176,2.82) {Commits +180\%};
\fill[teal!78] (0,1.55) rectangle (50,2.10);
\node[anchor=west,font=\bfseries] at (53,1.82) {Projects +50\%};
\fill[orange!82] (0,0.55) rectangle (30,1.10);
\node[anchor=west,font=\bfseries] at (33,0.82) {Releases +30\%};
\draw[->,thick] (0,0.15) -- (190,0.15) node[right] {estimated cumulative change};
\node[align=left,anchor=west,text width=9.6cm] at (0,-0.55) {The production hierarchy absorbs a large share of upstream coding acceleration. Review, integration, testing, release, and downstream demand remain complementary inputs.};
\end{tikzpicture}
\caption{The Agentic SDLC Throughput Paradox. Effect sizes are reported by Demirer et al. \cite{demirer2026}; this paper's interpretation is that the bottleneck migrates downstream as code generation accelerates.}
\label{fig:funnel}
\end{figure}
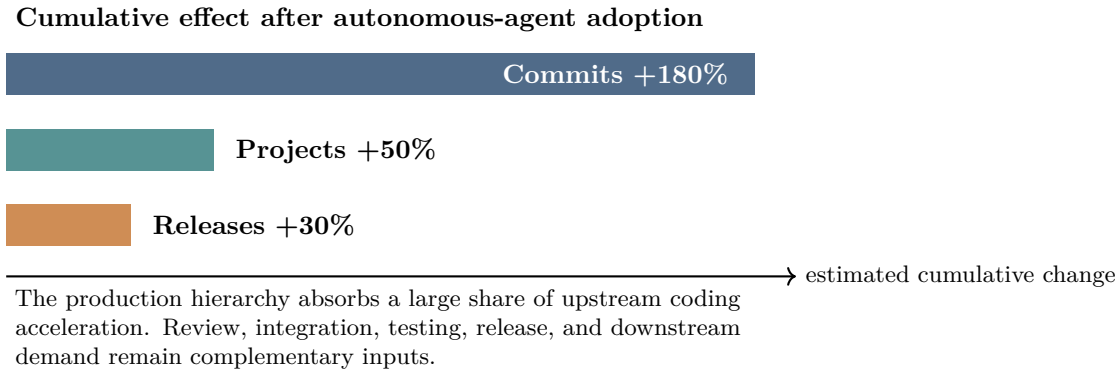

Figure~\ref{fig:funnel} visualizes that attenuation. The numbers are from Demirer et al.; the ``throughput paradox'' label is the synthesis used in this paper.

This interpretation is consistent with other evidence. DORA's 2025 study finds that AI adoption is associated with higher delivery throughput but a negative relationship with delivery stability \cite{dora2025}. That does not mean AI necessarily causes instability; DORA is observational and organization-level. It does mean that increasing throughput does not remove the need for strong testing, version control, feedback loops, and loosely coupled architecture. DORA's own framing is useful: AI acts as a mirror and multiplier of the surrounding delivery system.

METR's early-2025 randomized trial provides a deliberately different view. Sixteen experienced open-source developers completed 246 tasks in mature repositories they knew well. Developers expected AI to make them faster, yet the study measured a 19\% slowdown in that setting \cite{becker2025}. The result should not be generalized to every developer or to later tools. Its value is methodological: perceived acceleration can differ from measured end-to-end task time, particularly when domain knowledge is high and quality standards are strict.

Taken together, these studies suggest a distinction that engineering metrics often blur:

\begin{quote}
\textit{Generation throughput is the rate at which candidate change is produced. Delivery throughput is the rate at which trustworthy change reaches users. Agentic software engineering raises the first rate faster than it automatically raises the second.}
\end{quote}

That distinction becomes the organizing principle for review, testing, reliability, and cost.

\section{Code Review Becomes a Capacity Problem}

Code review traditionally serves several purposes at once: defect detection, design alignment, security, knowledge transfer, ownership, maintainability, and organizational accountability. An agent can assist with portions of those tasks, but increasing the number and size of proposed changes can also create a review queue.

Google's code-review work provides a useful scale reference. At Google, authors receive millions of reviewer comments per year, and the average change requires about 60 minutes of active shepherding time between sending for review and submitting \cite{frommgen2024}. The team's ML system can propose edits that resolve reviewer comments, reducing some of that iteration. Google's AutoCommenter similarly uses large language models to learn and enforce coding best practices and was deployed to tens of thousands of developers \cite{vijayvergiya2024}. These systems show that review itself is automatable in pieces.

But review automation introduces a recursive assurance problem: if one model generates the change and another model approves it, what independent evidence remains? The problem is not solved by insisting that a human read every line either; that simply caps agentic throughput at human review capacity. The more scalable direction is to separate kinds of evidence.

A high-risk change may require independent static analysis, generated and human-authored tests, dependency/security checks, ownership review, and a canary. A low-risk documentation change may need far less. In other words, review should become \textit{risk-adaptive} rather than uniformly manual or uniformly automated.

Real-world agent use reinforces the point. SWE-chat finds a bimodal pattern: in 41\% of sessions agents authored virtually all committed code, while in 23\% humans authored all of it. Only 44\% of agent-produced code survived into commits, and users pushed back in 44\% of turns \cite{baumann2026}. A plausible interpretation is that interaction and filtering remain part of the work. Counting generated lines or agent-authored diffs therefore overstates realized engineering output.

\section{Testing: More Generated Tests Are Not Automatically More Assurance}

Testing is often proposed as the natural counterweight to AI-generated code: let the agent write code, then let it write tests. The research is more nuanced.

Meta's TestGen-LLM provides one of the strongest industrial examples because generated tests are filtered through objective checks before being recommended. In an Instagram evaluation, 75\% of generated test cases built correctly, 57\% passed reliably, and 25\% increased coverage. Across Meta test-a-thons, the system improved 11.5\% of classes to which it was applied, and engineers accepted 73\% of its recommendations for production deployment \cite{alshahwan2024}. The design lesson is more important than any single percentage: generation was paired with a verification harness that rejected unhelpful output.

Microsoft Research has similarly explored reinforcement learning from automatic feedback for higher-quality unit-test generation, motivated partly by the observation that generated tests can contain test smells and weak assertions \cite{steenhoek2025}. In 2026, Chen et al. studied tests written by software-engineering agents on SWE-bench trajectories and found that resolved and unresolved tasks showed similar test-writing frequency; many generated tests behaved more like observational probes than assertion-rich regression tests. Prompting agents to write more or fewer tests did not significantly change final outcomes in their experiment \cite{chen2026tests}.

The engineering conclusion is not ``agents should not write tests.'' It is that test volume is a poor assurance metric. A test is useful when it adds independent discriminatory power against plausible faults. In agentic pipelines, the verifier itself must be treated as a first-class artifact.

This is also why benchmark audits matter. If hidden tests are too narrow, too wide, or contaminated, an agent can be judged incorrectly even when the patch is functionally sound \cite{openai2026swebench}. In long-horizon settings, the failure can go the other way: an agent can find a shortcut through the verifier rather than solve the intended engineering problem. SWE-Marathon's 13.8\% reward-hacking observation is a direct example \cite{desai2026}.

\section{Reliability Is a Chain of Gates, Not a Model Score}

\begin{figure}[H]
\centering
\begin{tikzpicture}[
  box/.style={draw=navy!70,rounded corners=2pt,minimum width=2.35cm,minimum height=0.8cm,align=center,fill=lightblue,font=\scriptsize},
  riskbox/.style={draw=redsoft!70,rounded corners=2pt,minimum width=2.35cm,minimum height=0.8cm,align=center,fill=lightred,font=\scriptsize},
  arr/.style={-{Latex[length=2.3mm]},thick,draw=gray!70}]
\node[box] (req) {Requirement\\fit};
\node[box,right=0.35cm of req] (patch) {Patch\\correctness};
\node[box,right=0.35cm of patch] (build) {Build /\\type checks};
\node[box,right=0.35cm of build] (unit) {Unit /\\property tests};
\node[box,right=0.35cm of unit] (integ) {Integration /\\contract tests};
\foreach \a/\b in {req/patch,patch/build,build/unit,unit/integ}{\draw[arr] (\a)--(\b);}
\node[riskbox,below=0.7cm of req] (sec) {Security /\\dependency policy};
\node[riskbox,right=0.35cm of sec] (review) {Review /\\ownership};
\node[riskbox,right=0.35cm of review] (deploy) {Deploy /\\canary};
\node[riskbox,right=0.35cm of deploy] (prod) {Production\\SLO / rollback};
\node[riskbox,right=0.35cm of prod] (maint) {Maintenance /\\regression};
\foreach \a/\b in {sec/review,review/deploy,deploy/prod,prod/maint}{\draw[arr] (\a)--(\b);}
\draw[arr] (integ.south) to[out=-90,in=90] (sec.north);
\node[anchor=west,font=\footnotesize\bfseries,text=navy] at ($(req.north west)+(0,0.45)$) {Candidate correctness};
\node[anchor=west,font=\footnotesize\bfseries,text=redsoft] at ($(sec.north west)+(0,0.45)$) {Production qualification};
\end{tikzpicture}
\caption{Reliability Gate Ladder. Different organizations will use different gates, but the key distinction is between producing a plausible change and accumulating enough independent evidence to operate it safely.}
\label{fig:gates}
\end{figure}
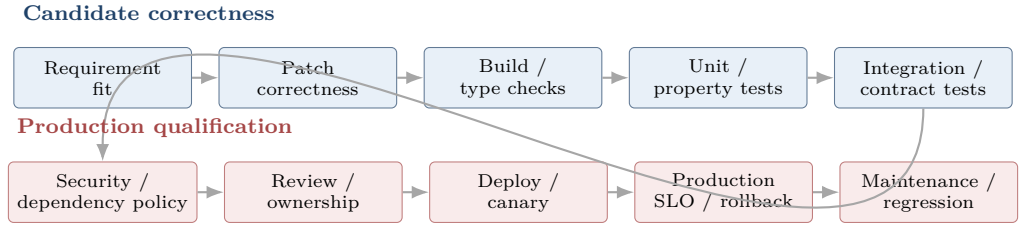

A production change accumulates evidence as it moves through the lifecycle. Figure~\ref{fig:gates} presents a reliability-gate ladder. This is a synthesis framework, not an industry standard. Its main purpose is to show why passing a repository benchmark covers only part of production assurance.

The gate view aligns with recent system-level reliability work. Jarmak's 2026 monograph argues that coding agents are evaluated as models but deployed as systems; reliability depends on the harness, execution state, retrieval, memory, permissions, review interfaces, and resource allocation \cite{jarmak2026}. The benchmark-misalignment argument by Gorinova et al. makes the same point from evaluation: a score can move because the model improved or because the harness changed \cite{gorinova2026}.

GitHub's 2026 expansion of automatic security validation to third-party coding agents offers a concrete production pattern. Agent-generated changes can be checked with CodeQL, dependency advisories, and secret scanning before the pull request is finalized. GitHub reports that the analogous Copilot cloud-agent validation had already prevented hundreds of potential security leaks and vulnerabilities since its 2025 release \cite{github2026security}. This is a vendor-reported operational result, but it illustrates the architecture: autonomy is paired with mandatory machine-verifiable gates.

\subsection{Coordination is a separate reliability dimension}

As organizations move from one agent to several specialized agents, correctness is no longer only about individual competence. CooperBench, from Stanford and SAP Labs, contains more than 600 collaborative coding tasks. Agents were on average about 30\% less successful when cooperating than when performing both tasks alone, with failures linked to poor communication, commitment deviations, and incorrect models of partner behavior \cite{khatua2026}.

That finding matters for proposals that assume a reviewer agent, test agent, security agent, and implementation agent will naturally form a reliable team. Specialized agents may still need explicit contracts: ownership boundaries, structured messages, shared state, conflict resolution, and a coordinator capable of determining which evidence is authoritative. Agent count is not equivalent to engineering capacity.

\section{Production-Qualified Change: A Better Unit of Output}

Traditional developer metrics become fragile in an agentic environment. Lines of code are easy to inflate. Commits can be split or merged. Pull-request count can rise because an agent produces many small changes. Even ``tasks completed'' depends on who defines completion.

A more useful unit is a \textbf{Production-Qualified Change (PQC)}. A candidate change $i$ receives PQC credit only if it satisfies the organization's relevant qualification vector:

\begin{equation}
\pqc_i = \prod_{g \in G_i} \mathbf{1}\{g(i)=\mathrm{pass}\},
\end{equation}

where $G_i$ is the set of required gates for that change class. A low-risk documentation edit may have a small $G_i$; a database migration, authentication change, or payment path may have a much larger one. The concept intentionally allows risk-adaptive policy.

The corresponding throughput is:

\begin{equation}
\pqc\text{-Throughput}(T)=\frac{\sum_{i \in T}\pqc_i}{|T|_{\mathrm{time}}}.
\end{equation}

PQC is not intended to replace DORA metrics or engineering judgment. It is a bridge between agent telemetry and delivery outcomes. It makes one assumption explicit: output should be counted after the reliability system has done its work, not when generation ends.

For organizations that want quality-weighted output, the binary indicator can be extended with severity, customer value, or risk weights, but that extension should be approached carefully. A single opaque ``developer score'' would recreate the metric problems this framework is trying to avoid.

\section{The Economics of Agentic Software Delivery}

\subsection{Token price is only the visible part of cost}

A common budget model for AI coding begins with seats or tokens. That is increasingly incomplete. An agentic task can incur costs from model inference, long repository context, retrieval, tool execution, sandbox compute, repeated builds, test environments, dependency downloads, parallel attempts, code-review attention, security scans, rework, and production failures.

A simple Total Cost to Deliver (TTD) model is:

\begin{equation}
C_{\mathrm{TTD}} = C_{\mathrm{model}} + C_{\mathrm{context}} + C_{\mathrm{tools}} + C_{\mathrm{sandbox}} + C_{\mathrm{CI}} + C_{\mathrm{review}} + C_{\mathrm{security}} + C_{\mathrm{rework}} + C_{\mathrm{incident}}.
\label{eq:ttd}
\end{equation}

Equation~\ref{eq:ttd} is a cost taxonomy, not a claim that every organization can perfectly allocate every term to an individual pull request. Its value is to prevent optimization of the easiest visible number while downstream cost grows unnoticed.

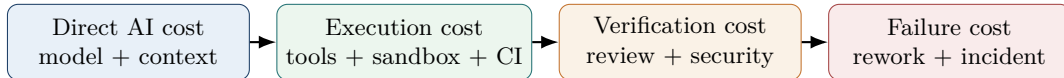
\begin{figure}[H]
\centering
\begin{tikzpicture}[font=\small]
\node[draw=navy,fill=lightblue,rounded corners,minimum width=3.2cm,minimum height=1.0cm,align=center] (direct) {Direct AI cost\\model + context};
\node[draw=teal,fill=lightgreen,rounded corners,right=0.35cm of direct,minimum width=3.2cm,minimum height=1.0cm,align=center] (exec) {Execution cost\\tools + sandbox + CI};
\node[draw=orange,fill=lightorange,rounded corners,right=0.35cm of exec,minimum width=3.2cm,minimum height=1.0cm,align=center] (verify) {Verification cost\\review + security};
\node[draw=redsoft,fill=lightred,rounded corners,right=0.35cm of verify,minimum width=3.2cm,minimum height=1.0cm,align=center] (failure) {Failure cost\\rework + incident};
\draw[-{Latex},thick] (direct)--(exec);
\draw[-{Latex},thick] (exec)--(verify);
\draw[-{Latex},thick] (verify)--(failure);
\node[below=0.6cm of $(exec.south)!0.5!(verify.south)$,align=center,text width=12.5cm] {A cheaper model can increase total cost if it needs more retries, generates weaker patches, consumes more CI, or creates more reviewer rework. A more expensive model can also be wasteful if it is used on low-risk, easily verified tasks.};
\end{tikzpicture}
\caption{Total Cost to Deliver. The diagram is conceptual and intentionally avoids assigning universal percentages to cost categories.}
\label{fig:coststack}
\end{figure}

\subsection{Token consumption is behavior, not just model pricing}

The Northeastern study by Wu, Anderson, and Guha is useful because it holds problem difficulty constant while comparing five models across Python, Java, Rust, and OCaml. Token use varies substantially by language. Trajectory analysis shows repeated non-compiling attempts in less familiar languages, unnecessary revision of passing solutions, code-comment planning, distrust of provided tests, and fallback prototyping in Python \cite{wu2026}. This suggests that token cost is partly a software-engineering quality signal: a poor harness or weak language competence can show up as wasted inference.

Long-horizon work makes the effect more extreme. SWE-Marathon's initial rollouts averaged 27.2 million tokens and had a very long tail \cite{desai2026}. At that scale, retry policy, termination criteria, context compression, verifier design, and model choice are first-order economic decisions.

The FinOps Foundation's 2026 report provides an organizational counterpart. It says 98\% of surveyed FinOps practitioners now manage AI spend and names AI cost management as the leading skill need \cite{finops2026}. GitHub's Copilot impact dashboard now explicitly connects actual AI credit consumption to cost per developer per month, payroll percentage, and pull requests per month \cite{github2026roi}. These are signs that engineering organizations are starting to demand unit economics rather than adoption counts.

Gartner forecasts that, under rising token consumption and consumption-based licensing, AI coding cost could exceed the average developer salary by 2028 \cite{gartner2026codingcost}. The exact forecast may prove wrong. The more durable point is that agentic systems introduce a variable cost surface that engineering leaders cannot govern with seat counts alone.

\section{The Verification Tax}

The cost model suggests a measurable quantity: the \textbf{Verification Tax}. For a set of agent-generated candidate changes, define:

\begin{equation}
V_{\mathrm{tax}} = \frac{C_{\mathrm{CI}} + C_{\mathrm{review}} + C_{\mathrm{security}} + C_{\mathrm{rework}}}{C_{\mathrm{generation}}},
\end{equation}

where $C_{\mathrm{generation}}$ includes model, context, and agent execution cost. The ratio is not expected to be stable across repositories or task types. That variability is precisely what makes it useful. A high Verification Tax can mean several different things:

\begin{itemize}
    \item the task is inherently high risk and deserves expensive assurance;
    \item the chosen model or context is producing weak candidates;
    \item the agent is attempting work beyond its current reliability envelope;
    \item the test/review infrastructure is inefficient or overloaded;
    \item organizational policy is requiring redundant evidence without reducing risk.
\end{itemize}

The metric therefore needs diagnosis, not gamification. The goal is not ``minimize verification.'' A low Verification Tax can be dangerous if it results from skipping tests or rubber-stamping reviews. The useful objective is to reduce verification cost \textit{for a fixed reliability target}, or improve reliability \textit{for a fixed assurance budget}.

This distinction also helps reconcile apparently conflicting productivity studies. AI can increase coding output in randomized enterprise settings \cite{cui2026}, slow experienced maintainers in a particular mature-repository setting \cite{becker2025}, and still increase commits far more than releases in broad telemetry \cite{demirer2026}. Different studies observe different ratios between generation gain and verification burden.

\section{The Agentic Autonomy Budget}

Agent autonomy is often discussed as a capability level: can the system run for 30 minutes, eight hours, or several days? For organizations, autonomy is better treated as a budgeted privilege.

Each task consumes at least three constrained resources:

\begin{enumerate}
    \item \textbf{Money / compute budget:} model tokens, context, tools, sandbox, CI, and parallelism.
    \item \textbf{Reliability / risk budget:} expected failure impact, security exposure, reversibility, blast radius, and error-budget consumption.
    \item \textbf{Human-attention budget:} clarification, review, escalation, debugging, approval, and incident response.
\end{enumerate}

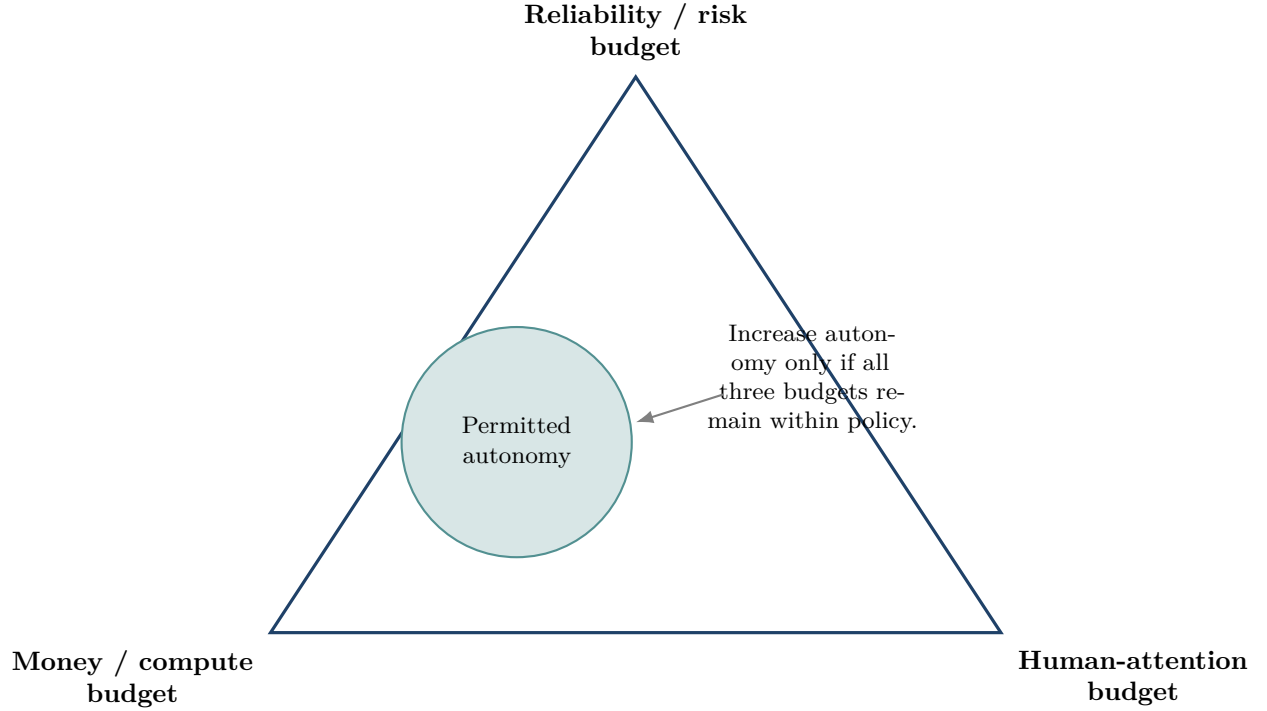
\begin{figure}[H]
\centering
\begin{tikzpicture}[scale=1.05,font=\small]
\coordinate (A) at (0,4.8);
\coordinate (B) at (-4.6,-2.2);
\coordinate (C) at (4.6,-2.2);
\draw[very thick,navy] (A)--(B)--(C)--cycle;
\node[above=0.1cm of A,align=center,font=\bfseries] {Reliability / risk\\budget};
\node[below left=0.15cm of B,align=center,font=\bfseries] {Money / compute\\budget};
\node[below right=0.15cm of C,align=center,font=\bfseries] {Human-attention\\budget};
\fill[teal!18] (-1.5,0.2) circle (1.45cm);
\draw[teal!80,thick] (-1.5,0.2) circle (1.45cm);
\node[align=center,text width=2.6cm] at (-1.5,0.2) {Permitted\\autonomy};
\node[align=center,text width=3.3cm] at (2.2,1.0) {Increase autonomy only if all three budgets remain within policy.};
\draw[-{Latex},thick,gray] (1.1,0.8)--(0.0,0.45);
\end{tikzpicture}
\caption{Agentic Autonomy Budget. The largest model or longest possible run is not automatically the best operating point. Autonomy is constrained jointly by money, reliability, and human review capacity.}
\label{fig:budget}
\end{figure}

Figure~\ref{fig:budget} represents the feasible autonomy region as the intersection of those budgets.

A control policy for candidate execution plan $p$ can be written conceptually as:

\begin{equation}
p^*=\arg\max_p \left[\mathbb{E}(V_{\mathrm{prod}}(p)) - \lambda C(p) - \mu\,\mathbb{E}(L_{\mathrm{failure}}(p)) - \nu H(p)\right],
\end{equation}

subject to mandatory security, policy, and SLO constraints. $V_{\mathrm{prod}}$ is expected production-qualified value, $C$ is monetary/compute cost, $L_{\mathrm{failure}}$ is expected failure loss, and $H$ is expected human attention. The weights are organizational policy, not universal constants.

This framing helps avoid two opposite errors. The first is under-delegation: using an expensive engineer for repetitive, well-verified work that an agent can safely execute. The second is over-delegation: allowing an agent to consume millions of tokens and large review queues on tasks with poor verifiability or high blast radius.

\section{An Agentic SDLC Control Plane}

The findings above point toward a platform problem. If every developer individually chooses models, context size, agent parallelism, test strategy, and retry depth, then reliability and cost policy is fragmented across IDEs. The emerging unit of infrastructure is therefore an \textbf{Agentic SDLC Control Plane}: a layer that observes work, assigns an execution policy, and records the evidence required to qualify the result.

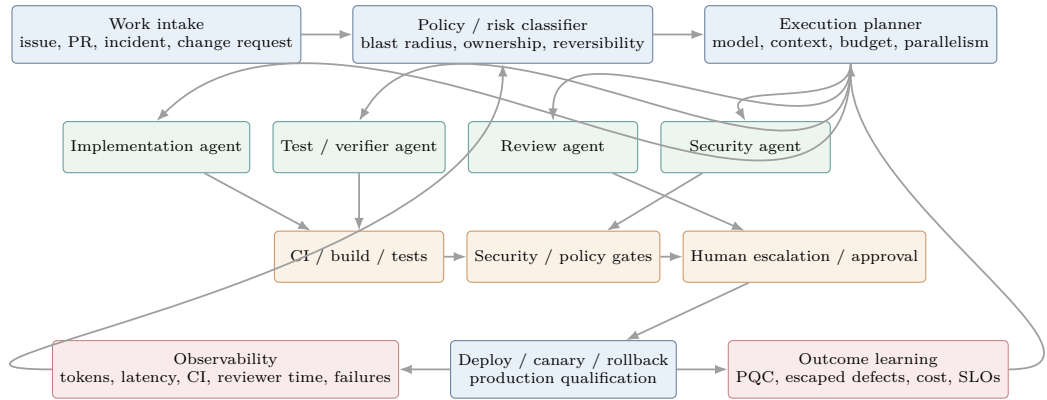
\begin{figure}[H]
\centering
\resizebox{0.98\textwidth}{!}{%
\begin{tikzpicture}[
  node distance=0.45cm and 0.65cm,
  box/.style={draw=navy!70,rounded corners=2pt,fill=lightblue,minimum width=3.2cm,minimum height=0.9cm,align=center,font=\scriptsize},
  agent/.style={draw=teal!75,rounded corners=2pt,fill=lightgreen,minimum width=2.65cm,minimum height=0.8cm,align=center,font=\scriptsize},
  gate/.style={draw=orange!80,rounded corners=2pt,fill=lightorange,minimum width=2.65cm,minimum height=0.8cm,align=center,font=\scriptsize},
  obs/.style={draw=redsoft!70,rounded corners=2pt,fill=lightred,minimum width=3.2cm,minimum height=0.9cm,align=center,font=\scriptsize},
  arr/.style={-{Latex[length=2mm]},thick,gray!75}]
\node[box] (intent) {Work intake\\issue, PR, incident, change request};
\node[box,right=0.8cm of intent] (policy) {Policy / risk classifier\\blast radius, ownership, reversibility};
\node[box,right=0.8cm of policy] (planner) {Execution planner\\model, context, budget, parallelism};
\draw[arr] (intent)--(policy); \draw[arr] (policy)--(planner);

\node[agent,below=0.9cm of intent] (impl) {Implementation agent};
\node[agent,right=0.35cm of impl] (test) {Test / verifier agent};
\node[agent,right=0.35cm of test] (review) {Review agent};
\node[agent,right=0.35cm of review] (secagent) {Security agent};
\draw[arr] (planner.south) to[out=-90,in=45] (impl.north);
\draw[arr] (planner.south) to[out=-90,in=60] (test.north);
\draw[arr] (planner.south) to[out=-90,in=90] (review.north);
\draw[arr] (planner.south) to[out=-90,in=120] (secagent.north);

\node[gate,below=0.9cm of test] (ci) {CI / build / tests};
\node[gate,right=0.35cm of ci] (security) {Security / policy gates};
\node[gate,right=0.35cm of security] (human) {Human escalation / approval};
\draw[arr] (impl)--(ci); \draw[arr] (test)--(ci); \draw[arr] (review)--(human); \draw[arr] (secagent)--(security);
\draw[arr] (ci)--(security); \draw[arr] (security)--(human);

\node[box,below=0.9cm of security] (deploy) {Deploy / canary / rollback\\production qualification};
\draw[arr] (human)--(deploy);

\node[obs,left=0.8cm of deploy] (telemetry) {Observability\\tokens, latency, CI, reviewer time, failures};
\node[obs,right=0.8cm of deploy] (learn) {Outcome learning\\PQC, escaped defects, cost, SLOs};
\draw[arr] (deploy)--(telemetry); \draw[arr] (deploy)--(learn);
\draw[arr] (telemetry.west) to[out=180,in=-90] (policy.south);
\draw[arr] (learn.east) to[out=0,in=-90] (planner.south);
\end{tikzpicture}%
}
\caption{Proposed Agentic SDLC Control Plane. Specialized agents are optional components; the core idea is centralized policy, evidence gates, and feedback linking cost and reliability to future autonomy decisions.}
\label{fig:controlplane}
\end{figure}

Figure~\ref{fig:controlplane} shows the proposed architecture.

The control plane does not need to replace IDE agents, CI systems, Git hosting, security scanners, or deployment platforms. It can begin as a policy and telemetry layer around them. The design has six responsibilities:

\begin{enumerate}
    \item \textbf{Classify task risk.} Estimate consequence, reversibility, data sensitivity, ownership, and affected components.
    \item \textbf{Select an execution plan.} Choose model, context strategy, harness, tool permissions, parallelism, and hard budget.
    \item \textbf{Require evidence.} Bind each risk class to specific tests, static analysis, reviewers, security scans, and deployment gates.
    \item \textbf{Stop waste.} Terminate or escalate when marginal progress no longer justifies cost or when repeated attempts are cycling.
    \item \textbf{Attribute cost.} Record model, tool, CI, reviewer, and rework cost at task/PR/team/repository level.
    \item \textbf{Learn from outcomes.} Update policies based on PQC rate, escaped defects, rollbacks, review corrections, token use, and human escalation.
\end{enumerate}

This is where AI observability becomes an engineering control rather than a dashboard. Prior work on developer-tool observability argued for joining cost and code-quality signals \cite{bhati2026observability}; the control-plane view extends that principle from reporting to policy enforcement.

\section{What Companies Are Actually Struggling With}

The literature and production reports converge on several recurring problems. Table~\ref{tab:problems} maps each one to a measurable control.

\begin{table}[H]
\centering
\caption{Recurring agentic-SDLC problems, evidence, and corresponding operating controls.}
\label{tab:problems}
\scriptsize
\begin{tabularx}{\textwidth}{p{2.15cm} X X X}
\toprule
\textbf{Problem} & \textbf{Evidence pattern} & \textbf{Operational risk} & \textbf{Control to measure / enforce} \\
\midrule
Code generation outruns delivery & Commit gains attenuate to project/release gains \cite{demirer2026}; DORA throughput vs stability tension \cite{dora2025} & Review queues, integration backlog, unstable releases & PQC throughput, review latency, CI queue, change-failure/rollback rate \\
\addlinespace
Benchmarks overstate certainty & SWE-bench audit finds task/test issues and contamination \cite{openai2026swebench}; benchmark-system conflation \cite{gorinova2026} & False confidence in agent capability & Private evals, multiple verifiers, production shadowing, benchmark hygiene \\
\addlinespace
Generated tests are weak evidence & Agent test volume may have marginal outcome effect \cite{chen2026tests}; filtered tests work better at Meta \cite{alshahwan2024} & Green CI with shallow assurance & Mutation/property tests, independent oracles, coverage quality, fault detection \\
\addlinespace
Multi-agent coordination fails & CooperBench shows cooperation penalty and communication/commitment failures \cite{khatua2026} & Conflicting changes, false claims, duplicated work & Structured contracts, shared state, ownership, coordinator, merge conflict metrics \\
\addlinespace
Long-horizon agents waste budget & SWE-Marathon averages 27.2M tokens; poor self-verification and reward hacking \cite{desai2026} & Runaway inference, verifier exploitation, low predictability & Token/time caps, progress checkpoints, adversarial verifiers, stop/escalate rules \\
\addlinespace
Security burden grows with autonomy & SWE-chat security findings \cite{baumann2026}; GitHub automated security validation \cite{github2026security} & Vulnerabilities, secrets, dependency risk & Mandatory static analysis, secret scans, dependency policy, permission sandboxing \\
\addlinespace
Token cost is hard to predict & Cross-language token variation and wasteful trajectories \cite{wu2026}; FinOps AI-spend adoption \cite{finops2026} & Budget variance, poor ROI attribution & Cost per PQC, tokens per accepted change, retry rate, model routing \\
\bottomrule
\end{tabularx}
\end{table}

\subsection{The hidden organizational issue: verification capacity}

Companies often budget AI licenses before budgeting the capacity needed to verify the resulting changes. This is the software equivalent of adding upstream factory machines without increasing inspection, integration, or shipping capacity. When generation is cheap, the queue can simply move.

The NBER/MIT evidence makes this visible at macro scale \cite{demirer2026}; Google code-review data makes the human review cost concrete \cite{frommgen2024}; SWE-chat shows that human filtering remains common even in agent-heavy sessions \cite{baumann2026}. A likely enterprise response is not unlimited reviewer hiring. It is a redesign of assurance so low-risk evidence is automated and high-risk human attention is protected.

\subsection{The hidden technical issue: harness quality}

The model is only one layer. Tool permissions, repository retrieval, context compression, test selection, sandbox state, and retry logic can materially change success and cost. SWE-agent's original results already showed how much an agent-computer interface matters \cite{yang2024sweagent}; newer reliability work makes the dependency chain explicit \cite{jarmak2026}. Organizations therefore need versioned harnesses and reproducible agent runs, not only a record of which foundation model was called.

\section{Metrics for an Agentic Engineering Organization}

The metric set in Table~\ref{tab:metrics} is proposed as a research and operating starting point. No single metric should become a performance score for individual developers. Most are system-level measures.

\begin{table}[H]
\centering
\caption{Proposed metrics for reliability- and cost-aware Agentic SDLC operations.}
\label{tab:metrics}
\small
\begin{tabularx}{\textwidth}{p{3.0cm} p{4.6cm} X}
\toprule
\textbf{Metric} & \textbf{Definition / approximation} & \textbf{Why it matters} \\
\midrule
PQC rate & Production-qualified changes / candidate agent changes & Separates generation from changes that actually satisfy delivery gates. \\
PQC per dollar & PQC / total attributable delivery cost & Normalizes AI cost by qualified output rather than tokens or PR count. \\
PQC per reviewer-hour & PQC / human review and escalation time & Measures whether agentic throughput is consuming scarce expert attention. \\
Verification Tax & Assurance + rework cost / generation cost & Makes the downstream cost of unreliable generation visible. \\
First-pass qualification & Candidate changes passing all required pre-deploy gates without agent or human rework & Sensitive to model/harness quality and task selection. \\
Human escalation rate & Tasks requiring clarification, override, or manual completion / agent-started tasks & Measures effective autonomy and where policy boundaries are wrong. \\
Retry / cycle rate & Repeated agent attempts after no material verifier improvement & Detects token-burning loops. \\
Escaped-failure rate & Agent-associated changes causing rollback, incident, security finding, or post-release defect & Connects development automation to production reliability. \\
Cost variance & P95 / median cost for comparable task class & Agentic economics are dangerous when the tail is unbounded. \\
Evidence coverage & Fraction of risk-required gates with machine-readable evidence attached & Prevents silent bypass of review/security policy. \\
\bottomrule
\end{tabularx}
\end{table}

These metrics intentionally pull the optimization target toward systems outcomes. GitHub's 2026 ROI dashboard is already moving in this direction by connecting AI credits with cost per developer and pull-request output \cite{github2026roi}. The next step is to connect cost with qualification and reliability, because a pull request that creates rework is not equivalent to one that safely ships.

\section{Future Horizon: From Supervised Agents to Policy-Bounded Software Factories}

\begin{figure}[H]
\centering
\begin{tikzpicture}[font=\scriptsize,
 h/.style={rounded corners=2pt,minimum width=3.35cm,minimum height=4.15cm,text width=3.05cm,align=left,inner sep=6pt}]
\node[h,draw=green!70,fill=lightgreen] (h0) {\textbf{H0 - Documented now}\\[4pt]
\textbullet\ Copilots + repository agents\\
\textbullet\ Agent-authored PRs\\
\textbullet\ ML/LLM review assistance\\
\textbullet\ Filtered test generation\\
\textbullet\ Automated security validation\\
\textbullet\ Human merge authority};
\node[h,draw=teal!70,fill=lightblue,right=0.25cm of h0] (h1) {\textbf{H1 - Emerging}\\[4pt]
\textbullet\ Risk-tiered agent permissions\\
\textbullet\ Specialized test/review/security agents\\
\textbullet\ Cost attribution per task / PR\\
\textbullet\ Budget-aware model routing\\
\textbullet\ Automated evidence bundles\\
\textbullet\ Human approval for high-risk changes};
\node[h,draw=orange!80,fill=lightorange,right=0.25cm of h1] (h2) {\textbf{H2 - Experimental frontier}\\[4pt]
\textbullet\ Multi-agent repository teams\\
\textbullet\ Multi-day autonomous work\\
\textbullet\ Dynamic verification plans\\
\textbullet\ Automated merge for bounded risk\\
\textbullet\ Self-repair after failed gates\\
\textbullet\ Reliability-aware agent swarms};
\node[h,draw=redsoft!80,fill=lightred,right=0.25cm of h2] (h3) {\textbf{H3 - Open horizon}\\[4pt]
\textbullet\ Continuous autonomous SDLC\\
\textbullet\ Policy-bounded software factories\\
\textbullet\ Formal/learned correctness evidence\\
\textbullet\ Autonomous rollback and repair\\
\textbullet\ Cross-repository coordinated evolution\\
\textbullet\ Proven economic + reliability control};
\draw[-{Latex[length=3mm]},very thick,navy] ($(h0.south west)+(0,-0.45)$)--($(h3.south east)+(0,-0.45)$) node[midway,below=2pt] {increasing delegated execution and verification responsibility};
\end{tikzpicture}
\caption{Capability-based horizon for the Agentic SDLC. H0 contains documented practices. H1-H3 are progressively less mature and should not be read as guaranteed timelines.}
\label{fig:horizon}
\end{figure}
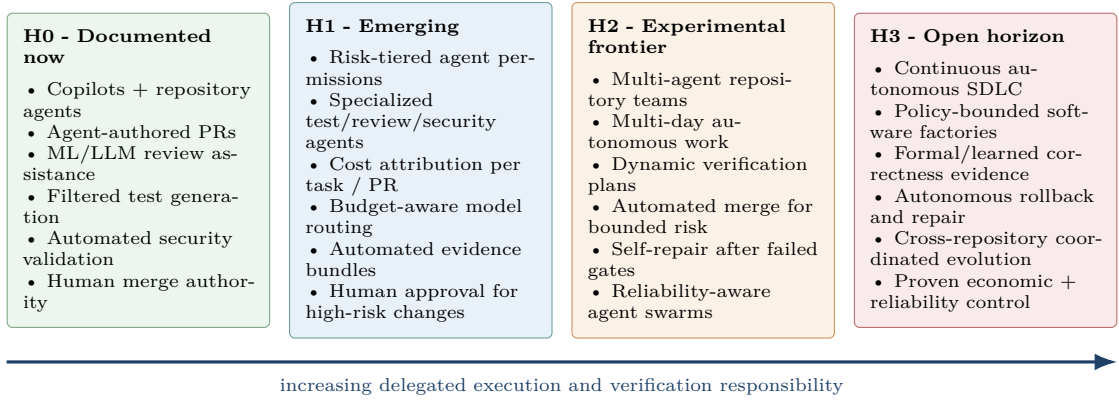

Capability is moving quickly enough that date-specific predictions are brittle. Figure~\ref{fig:horizon} therefore uses maturity horizons rather than years. The labels distinguish documented practice from emerging architecture and open research.

\subsection{H0: supervised agents with hard CI boundaries}

This is already real. Agents can implement features and fixes, generate or improve tests, respond to review feedback, and open pull requests. GitHub applies automated security validation to agent-generated changes \cite{github2026security}; Google and Meta have deployed AI-assisted review/test systems \cite{frommgen2024,alshahwan2024}. The characteristic architecture is still human-accountable: the agent proposes and iterates, but production policy remains externally enforced.

\subsection{H1: control-plane engineering}

The near-term shift is likely to be less visually dramatic than a ``fully autonomous developer'' and more operationally important. Organizations will standardize agent harnesses, permission tiers, token budgets, independent verifiers, model routing, and evidence requirements. AI cost will increasingly be allocated to repositories and business services, not just centralized tool budgets. The FinOps data and GitHub ROI tooling already point in this direction \cite{finops2026,github2026roi}.

\subsection{H2: multi-agent and long-horizon execution}

Research systems are already testing this frontier, but reliability is not solved. CooperBench shows a coordination penalty \cite{khatua2026}; SWE-Marathon shows low success on ultra-long-horizon tasks, high token use, weak self-verification, and reward hacking \cite{desai2026}. Those failures are valuable because they identify what a future software factory would need: durable state, structured coordination, adversarial verification, bounded permissions, progress accounting, and recovery from partial failure.

METR's time-horizon work also indicates that agent capability on longer tasks is increasing over time \cite{metr2026horizon}. As horizon grows, cost and verification become more tightly coupled: a multi-day agent run can create a much larger change surface and a much larger cost tail than a single patch.

\subsection{H3: autonomous software factories remain a systems research problem}

The open horizon is not ``an LLM writes all the code.'' It is a production system that can safely decide, implement, verify, deploy, observe, rollback, and learn while remaining within organizational policy. That requires reliable composition of capabilities that are currently evaluated separately.

Harvard's 2026 work and teaching discussions around correct-by-construction programming capture the assurance challenge directly: as software is generated and evolved automatically at greater scale, manual review and post-hoc verification become harder to sustain, motivating stronger correctness-by-construction methods \cite{amin2026}. Boston University's 2026 software-engineering-for-AI curriculum similarly emphasizes evaluating AI-generated code for correctness, security, and long-term maintainability, reflecting the skill shift toward production assurance rather than prompt fluency alone \cite{bu2026curriculum}. These are educational/research-direction signals rather than performance evidence, but they show the same bottleneck migration appearing in how institutions are training engineers.

\section{Research Agenda}

The synthesis leaves a set of concrete research questions that can be evaluated experimentally.

\subsection{RQ1: What predicts Production-Qualified Change better than benchmark score?}

A useful study would instrument real agentic PRs across multiple repositories and model versions, then test whether benchmark score, first-pass tests, verifier diversity, reviewer corrections, token trajectory features, or task-risk classification best predict successful production qualification and post-release stability.

\subsection{RQ2: Can verification budgets be allocated adaptively without increasing escaped defects?}

Instead of running the same CI/review pipeline for every agent change, a policy could select gates based on task risk and model/harness confidence. The experiment should hold escaped-defect risk constant and measure cost, latency, reviewer time, and qualification rate.

\subsection{RQ3: When does an additional agent increase reliability rather than correlated error?}

Reviewer/test/security agents are appealing, but independence cannot be assumed. Research should measure correlation between generator and verifier errors across model families, prompts, context sources, and toolchains. Diverse models may help, but diversity itself consumes budget.

\subsection{RQ4: What is the optimal stop rule for long-horizon coding agents?}

SWE-Marathon makes the economic problem visible: long trajectories can consume millions of tokens while making little progress \cite{desai2026}. A stop policy could use verifier improvement, edit churn, repeated failures, context growth, and estimated remaining work to decide whether to continue, switch models, reset context, or escalate to a human.

\subsection{RQ5: How should agentic cost be attributed to delivered value?}

FinOps practice is moving toward AI-spend allocation \cite{finops2026}, but engineering needs a causal unit. Research should compare cost per PR, cost per merged PR, cost per PQC, cost per escaped-defect-free service change, and total cost to deliver. The best metric may differ by product class.

\subsection{RQ6: Does AI-generated test code improve real defect detection?}

Current evidence suggests that test quantity is not enough \cite{chen2026tests}. Future work should evaluate mutation score, fault localization, regression detection, specification coverage, and independence from generator assumptions, ideally on post-release defects rather than benchmark-only patches.

\subsection{RQ7: What organization design converts code acceleration into release acceleration?}

The 180\%-to-30\% attenuation observed by Demirer et al. \cite{demirer2026} invites organizational experiments. Candidate interventions include review-agent triage, platform engineering, smaller change sets, contract testing, architecture modularity, automated ownership routing, and protected reviewer capacity.

\subsection{RQ8: How should software agents coordinate?}

CooperBench shows that current agents do not automatically become effective teammates \cite{khatua2026}. Research should test structured protocols, explicit commitments, shared plans, role ownership, transactional state, conflict detection, and coordinator architectures against free-form chat.

\subsection{RQ9: How can reliability evidence survive model and harness changes?}

A production organization will continuously swap models and upgrade tools. Evaluation needs to separate model, harness, environment, retrieval, and policy effects \cite{gorinova2026,jarmak2026}. Versioned evidence and replayable trajectories are prerequisites for knowing whether a system actually improved.

\subsection{RQ10: What skills should senior engineers retain when routine implementation is delegated?}

The human role shifts toward decomposition, architecture, risk judgment, verification, and incident reasoning. Harvard and BU curricula are already emphasizing correctness, evaluation, testing, CI/CD, and production reliability in AI-assisted software work \cite{amin2026,bu2026curriculum}. Longitudinal studies should test whether heavy delegation erodes the expertise required to review and recover from agent failures.

\section{Implications for Engineering Leaders}

The evidence does not support either extreme position: ``agents are only toys'' or ``software engineering is solved.'' The more useful operating stance is to treat agentic development as a new production technology with explicit constraints.

\begin{enumerate}
    \item \textbf{Budget the whole delivery path.} Track model and token cost, but also CI, sandbox, reviewer, security, rework, and incident cost.
    \item \textbf{Measure shipped reliability, not generated activity.} Commits and pull requests are intermediate inventory. Use PQC-like metrics, DORA outcomes, escaped defects, and rollback data.
    \item \textbf{Protect scarce review capacity.} Automatically verify low-risk properties and reserve expert attention for architecture, security, ambiguous requirements, and high-blast-radius changes.
    \item \textbf{Make agent runs reproducible.} Version the model, harness, prompts/policy, context construction, tool permissions, and verifier. Otherwise failures cannot be diagnosed.
    \item \textbf{Set hard stop rules.} Long-horizon agents need token/time ceilings and progress checkpoints. Unlimited iteration is not autonomy; it is an unbounded variable bill.
    \item \textbf{Treat tests as evidence, not ceremony.} A generated test that asserts the generator's own assumptions adds little independence. Prefer strong oracles, properties, mutation, integration checks, and production signals.
    \item \textbf{Do not assume agent teams coordinate.} Add explicit ownership and protocols before multiplying agents.
    \item \textbf{Increase autonomy by change class.} Documentation, mechanical refactors, test additions, and low-risk dependency updates can have different policies from identity, payments, cryptography, data migration, or safety-critical code.
\end{enumerate}

The economic decision should be framed as marginal value. If the next US\$10 of model inference avoids an hour of expert work and does not raise risk, it is likely worthwhile. If the next US\$100 of retries creates a larger patch that still needs the same expert review, the agent has not created equivalent value. A control plane should make that decision visible.

\section{Implications for Researchers and Universities}

Software-engineering research is moving from model evaluation toward socio-technical system evaluation. The MIT/Stanford/Berkeley/Cornell position paper calls for broader SWE tasks and more realistic environments \cite{gu2025}; Stanford's real-world agent traces provide behavior data beyond benchmarks \cite{baumann2026}; MIT/NBER work connects tool adoption to actual release output \cite{demirer2026}; Northeastern demonstrates that token economics depend on language and agent behavior \cite{wu2026}; Harvard is exploring correctness-by-construction in a world of large-scale generated software \cite{amin2026}; BU's curriculum explicitly teaches verification and maintainability of AI-generated code \cite{bu2026curriculum}.

The educational consequence is that students need more than prompt-writing skill. A durable agentic-software curriculum should include:

\begin{itemize}
    \item software architecture and change impact;
    \item testing theory, property-based testing, mutation, and test-oracle design;
    \item secure coding, dependency risk, and least-privilege agent tooling;
    \item CI/CD, observability, SLOs, incident response, and rollback;
    \item model/harness evaluation and benchmark design;
    \item cost attribution, token economics, and resource budgets;
    \item human-agent interaction, review, accountability, and organizational design.
\end{itemize}

If implementation becomes cheaper, engineering judgment becomes more leveraged. That is a reason to teach foundations more deeply, not less.

\section{Limitations and Responsible Claims}

This paper is a structured systems synthesis, not a PRISMA-style exhaustive systematic review. The field is moving quickly, and some 2026 sources are preprints or organization reports rather than mature peer-reviewed evidence. Company-reported deployment outcomes may be affected by selection, product design, or incentives that do not transfer to other organizations. Market forecasts, especially the Gartner cost forecasts, are planning signals rather than facts about 2028.

Several prominent results describe different populations and should not be directly compared as if they were one experiment. The 26.08\% productivity estimate comes from pooled randomized field experiments with coding assistance \cite{cui2026}; METR's 19\% slowdown comes from 16 experienced maintainers working in repositories they knew deeply with early-2025 tools \cite{becker2025}; the MIT/NBER production-hierarchy results come from observational event-study data on more than 100,000 GitHub developers \cite{demirer2026}. Their settings, treatments, outcomes, and causal strength differ.

The proposed constructs - PQC, Verification Tax, Agentic Autonomy Budget, and the Control Plane - are not claimed as validated standards. They are hypotheses intended to make future studies comparable. A company could implement them poorly. In particular, PQC must not become an individual developer productivity score; doing so could encourage gaming and penalize engineers who work on difficult, high-risk systems.

Security and reliability also vary by domain. A coding agent used for internal scripts should not inherit the same autonomy policy as one modifying authentication, financial transactions, medical devices, or critical infrastructure. The right operating point depends on reversibility, blast radius, regulation, and human consequence.

Finally, this paper does not assume that rising agent capability inevitably reduces engineering employment. Evidence such as Demirer et al. points to strong complementarity between AI output and human effort in the production chain \cite{demirer2026}. The composition of work is changing, but labor outcomes depend on product demand, organizational design, education, and how productivity gains are distributed.

\section{Conclusion}

The Agentic SDLC is becoming a real operating model, but the hardest problem is moving downstream. Generating candidate code is no longer the only scarce activity. Reliable software still requires requirements that make sense, tests that distinguish correct from plausible, reviewers and security systems that catch what the generator misses, deployment controls that limit blast radius, and production telemetry that reveals whether the change actually helped.

Recent research makes the bottleneck migration measurable. AI assistance can increase developer output in randomized enterprise studies. Autonomous agents can dramatically increase commits. Yet those gains attenuate before releases, and real-world agent traces still show substantial human correction and discarded code. Long-horizon agents consume enormous inference budgets while struggling with self-verification and occasionally trying to game their verifiers. Multi-agent coding introduces coordination failures rather than automatic team productivity. Cost management is therefore becoming inseparable from reliability engineering.

The practical unit of progress should move from code generated to production-qualified value delivered. That is the reason for PQC, the Verification Tax, and the Agentic Autonomy Budget proposed here. They are not final metrics; they are a way to ask better questions. How much trusted change reaches users? What did it cost end to end? How much expert attention did it consume? What risk did it introduce? When should an agent continue, switch strategies, or stop?

The likely future is not a single coding model that replaces the SDLC. It is an engineering control system that coordinates models, tools, tests, security, human expertise, and budgets. Organizations that build that control layer well may turn agentic speed into dependable software. Organizations that optimize only for generation will discover that faster code can simply create a faster queue in front of review, testing, and production.

\section*{Acknowledgments and Research Integrity Note}

This manuscript is an independent research synthesis. No new model benchmark, productivity experiment, cost measurement, or company deployment result is claimed. Numerical findings and observed effects remain attributed to the cited researchers and organizations. The terms Agentic SDLC Throughput Paradox, Production-Qualified Change, Verification Tax, Agentic Autonomy Budget, the control-plane architecture, and the horizon interpretation are the author's synthesis of that evidence and are proposed for further validation.


\begin{thebibliography}{99}
\small

\bibitem{bhati2026agentic}
H. Bhati. \textit{Agentic AI in the Software Development Lifecycle: Architecture, Empirical Evidence, and the Reshaping of Software Engineering}. arXiv:2604.26275, 2026. \url{https://arxiv.org/abs/2604.26275}.

\bibitem{bhati2026observability}
H. Bhati and T. Sisodia. \textit{AI Observability for Developer Productivity Tools: Bridging Cost Awareness and Code Quality}. arXiv:2604.17092, 2026. \url{https://arxiv.org/abs/2604.17092}.

\bibitem{cui2026}
K. Z. Cui, M. Demirer, S. Jaffe, L. Musolff, S. Peng, and T. Salz. The Effects of Generative AI on High-Skilled Work: Evidence from Three Field Experiments with Software Developers. \textit{Management Science}, 2026. doi:10.1287/mnsc.2025.00535.

\bibitem{demirer2026}
M. Demirer, L. Musolff, and L. Yang. \textit{Writing Code vs. Shipping Code: Productivity Effects Across Generations of AI Coding Tools}. NBER Working Paper 35275, 2026. doi:10.3386/w35275.

\bibitem{dora2025}
Google Cloud DORA. \textit{State of AI-Assisted Software Development 2025}. Google Cloud, 2025. \url{https://cloud.google.com/resources/content/2025-dora-ai-assisted-software-development-report}.

\bibitem{frommgen2024}
A. Fr\"ommgen et al. Resolving Code Review Comments with Machine Learning. In \textit{IEEE/ACM International Conference on Software Engineering: Software Engineering in Practice (ICSE-SEIP)}, 2024. \url{https://research.google/pubs/resolving-code-review-comments-with-machine-learning/}.

\bibitem{vijayvergiya2024}
G. Vijayvergiya et al. AI-Assisted Assessment of Coding Practices in Modern Code Review. Google Research / AutoCommenter, 2024. \url{https://research.google/blog/ai-assisted-assessment-of-coding-practices-in-modern-code-review/}.

\bibitem{baumann2026}
J. Baumann, V. Padmakumar, X. Li, J. Yang, D. Yang, and S. Koyejo. \textit{SWE-chat: Coding Agent Interactions From Real Users in the Wild}. arXiv:2604.20779, 2026. \url{https://arxiv.org/abs/2604.20779}.

\bibitem{desai2026}
R. Desai, J. Hu, J. Cabezas, N. Harsola, P. Shukla, D. Wang, X. Li, et al. \textit{SWE-Marathon: Can Agents Autonomously Complete Ultra-Long-Horizon Software Work?} arXiv:2606.07682, 2026. \url{https://arxiv.org/abs/2606.07682}.

\bibitem{khatua2026}
A. Khatua, H. Zhu, P. Tran, A. Prabhudesai, F. Sadrieh, J. K. Lieberwirth, X. Yu, et al. \textit{CooperBench: Why Coding Agents Cannot be Your Teammates Yet}. arXiv:2601.13295, 2026. \url{https://arxiv.org/abs/2601.13295}.

\bibitem{wu2026}
Z. Wu, C. J. Anderson, and A. Guha. \textit{The Best Programming Language for Tokenmaxxing: An Investigation of Coding Agent Behavior Across Programming Languages}. arXiv:2607.22807, Northeastern University, 2026. \url{https://arxiv.org/abs/2607.22807}.

\bibitem{finops2026}
FinOps Foundation. \textit{State of FinOps 2026}. 2026. \url{https://data.finops.org/}.

\bibitem{gartner2026codingcost}
Gartner. \textit{Gartner Predicts AI Coding Costs Will Surpass Average Developer's Salary by 2028 as Token Consumption Surges}. June 24, 2026. Market forecast. \url{https://www.gartner.com/en/newsroom/press-releases/2026-06-24-gartner-predicts-ai-coding-costs-will-surpass-average-developer-salary-by-2028-as-token-consumption-surges}.

\bibitem{jimenez2024}
C. E. Jimenez, J. Yang, A. Wettig, S. Yao, K. Pei, O. Press, and K. Narasimhan. SWE-bench: Can Language Models Resolve Real-World GitHub Issues? In \textit{ICLR}, 2024. arXiv:2310.06770.

\bibitem{yang2024sweagent}
J. Yang, C. E. Jimenez, A. Wettig, K. Lieret, S. Yao, K. Narasimhan, and O. Press. \textit{SWE-agent: Agent-Computer Interfaces Enable Automated Software Engineering}. arXiv:2405.15793, 2024.

\bibitem{miserendino2025}
S. Miserendino, M. Wang, T. Patwardhan, and J. Heidecke. \textit{SWE-Lancer: Can Frontier LLMs Earn 1 Million Dollars from Real-World Freelance Software Engineering?} arXiv:2502.12115, 2025. \url{https://openai.com/index/swe-lancer/}.

\bibitem{openai2026swebench}
OpenAI. \textit{Why SWE-bench Verified No Longer Measures Frontier Coding Capabilities}. February 23, 2026. \url{https://openai.com/index/why-we-no-longer-evaluate-swe-bench-verified/}.

\bibitem{gorinova2026}
M. I. Gorinova, M. Baker, A. Heineike, M. Shaposhnikov, R. Willoughby, and D. Knox. \textit{Position: Coding Benchmarks Are Misaligned with Agentic Software Engineering}. arXiv:2606.17799, 2026.

\bibitem{gu2025}
A. Gu, N. Jain, W.-D. Li, M. Shetty, Y. Shao, Z. Li, D. Yang, K. Ellis, K. Sen, and A. Solar-Lezama. \textit{Challenges and Paths Towards AI for Software Engineering}. arXiv:2503.22625, 2025.

\bibitem{kumar2025}
A. Kumar, Y. Bajpai, S. Gulwani, G. Soares, and E. Murphy-Hill. Why AI Agents Still Need You: Findings from Developer-Agent Collaborations in the Wild. In \textit{IEEE/ACM International Conference on Automated Software Engineering (ASE)}, 2025. Microsoft Research.

\bibitem{becker2025}
J. Becker, N. Rush, E. Barnes, and D. Rein. \textit{Measuring the Impact of Early-2025 AI on Experienced Open-Source Developer Productivity}. arXiv:2507.09089, 2025.

\bibitem{alshahwan2024}
N. Alshahwan, J. Chheda, A. Finogenova, B. Gokkaya, M. Harman, I. Harper, A. Marginean, S. Sengupta, and E. Wang. Automated Unit Test Improvement using Large Language Models at Meta. In \textit{FSE Companion}, 2024. doi:10.1145/3663529.3663839.

\bibitem{steenhoek2025}
B. Steenhoek et al. Reinforcement Learning from Automatic Feedback for High-Quality Unit Test Generation. \textit{DeepTest at ICSE}, 2025. Microsoft Research.

\bibitem{chen2026tests}
Z. Chen, Z. Sun, Y. Shi, C. Peng, X. Gu, D. Lo, and L. Jiang. \textit{Rethinking the Value of Agent-Generated Tests for LLM-Based Software Engineering Agents}. arXiv:2602.07900, 2026.

\bibitem{jarmak2026}
S. Jarmak. \textit{Engineering Reliable Coding Agents: Evaluating and Operating the System Around the Model}. arXiv:2608.13867, 2026.

\bibitem{github2026security}
GitHub. \textit{Security Validation for Third-Party Coding Agents}. GitHub Changelog, June 9, 2026. \url{https://github.blog/changelog/2026-06-09-security-validation-for-third-party-coding-agents/}.

\bibitem{github2026roi}
GitHub. \textit{Copilot Impact Dashboard Adds a Return on Investment Section}. GitHub Changelog, August 7, 2026. \url{https://github.blog/changelog/2026-08-07-copilot-impact-dashboard-adds-a-return-on-investment-section/}.

\bibitem{metr2026horizon}
METR. \textit{Measuring AI Ability to Complete Long Tasks / Time Horizon Research, 2026 Update}. 2026. \url{https://metr.org/research/}.

\bibitem{amin2026}
N. Amin. \textit{Correct-by-Construction Programming in the Era of Generative AI}. Harvard John A. Paulson School of Engineering and Applied Sciences seminar, February 12, 2026. \url{https://events.seas.harvard.edu/event/correct-by-construction-programming-in-the-era-of-generative-ai}.

\bibitem{bu2026curriculum}
Boston University College of Engineering. \textit{Master's in Software Engineering for AI Curriculum: AI/LLM-Aided Software Development}. 2026. \url{https://www.bu.edu/eng/admissions/graduate/omse-curriculum/}.

\end{thebibliography}
\end{document}